\documentclass[aps,prl, preprint, amsmath, amssymb, superscriptaddress]{revtex4-2}
\usepackage{graphicx}
\usepackage{xcolor}
\begin{document}

\title{A new approach for solving the problem of creation of inverse electron distribution function and practical
recommendations for experimental searches for such media in glow discharges with hollow and flat cathodes}

\author{Chengxun Yuan}
\affiliation{School of Physics, Harbin Institute of Technology, Harbin 150001, China}
\affiliation{Heilongjiang Provincial Innovation Research Center for Plasma Physics and Application Technology}
\email{yuancx@hit.edu.cn}

\author{E.A. Bogdanov}
\affiliation{School of Physics, Harbin Institute of Technology, Harbin 150001, China}
\affiliation{Faculty of Physics, St. Petersburg State University, 198504 St. Petersburg, Russia}
\email{eugene72@mail.ru}

\author{A.A. Kudryavtsev}
\affiliation{School of Physics, Harbin Institute of Technology, Harbin 150001, China}
\affiliation{Heilongjiang Provincial Innovation Research Center for Plasma Physics and Application Technology}

\author{Jingfeng Yao}
\affiliation{School of Physics, Harbin Institute of Technology, Harbin 150001, China}
\affiliation{Heilongjiang Provincial Innovation Research Center for Plasma Physics and Application Technology}

\author{Zhongxiang Zhou}
\affiliation{School of Physics, Harbin Institute of Technology, Harbin 150001, China}
\affiliation{Heilongjiang Provincial Innovation Research Center for Plasma Physics and Application Technology}

\begin{abstract}
This paper proposes a novel approach for creating an inverse electron distribution function (EDF). Based on the obtained criteria for the formation of an inverse EDF in a non-uniform plasma, studies are conducted in low- and medium-pressure glow discharges with flat and hollow cathodes. The results of the numerical modeling and theoretical analysis are used to present reliable criteria and scaling for the evaluation of the possible inversion of the EDF under specific conditions. By solving the nonlocal Boltzmann kinetic equation in energy and coordinate variables, it is shown that the simplest way to implement the inversion of the EDF is in a glow discharge with a hollow cathode. For such discharges, practical recommendations are developed and specific conditions for the experimental detection of an inverse EDF are identified.
\end{abstract}

\maketitle

The search for nonequilibrium media with an inverse distribution of particles by energy states is crucial for the development of science and practice. Bekefi et al. \cite{bib01} showed that, almost immediately after the creation of the first gas lasers, a medium with inverse free electron distribution function (EDF) in gases with a Ramsauer minimum of the elastic scattering cross section will have an absolute negative conductivity of the electron gas, and thus it will amplify electromagnetic waves in a wide frequency range.
However, in contrast to the laser technology, based on the inversion of the bound electron states of atoms and molecules, this problem of creating an inverse EDF is not solved yet.

A new approach for creating an inverse EDF was developed in \cite{bib02, bib03, bib04}. It was demonstrated that a key reason for the current situation is that the search for an inverse EDF in the literature has typically involved a significant simplification of the Boltzmann kinetic equation, assuming a spatially homogeneous medium where the equation depends only on a single variable (the electron velocity) \cite{bib05}.

In practical applications, real laboratory plasma objects are always bounded and the EDF depends coordinates.
Consequently, this paper develops a novel approach for solving the problem of practical implementation of an inverse EDF.

In \cite{bib02, bib03}, it was first shown that for the formation of the inversion of the EDF $f_0$
at low kinetic energies $w$, where the EDF reaches its maximum value, the boundary condition
\begin{equation}\label{eq01}
{\left. {\left( {\nabla \varphi \, \cdot \,\nabla {f_0}} \right)} \right|_{w = 0}} < 0
\end{equation}
should be satisfied, where $\varphi$ is the electric potential.
Condition \eqref{eq01} can be satisfied only in a non-uniform plasma. For example when electric field accelerates electrons
to the anode but the EDF decreases in this direction. Also it can be observed that \eqref{eq01} is inconsistent with the traditional Boltzmann distribution
$n_e \sim \text{exp}(\varphi / T_e)$ ($n_e$ and $T_e$ are the electron density and electron temperature, respectively).

Thus, when searching for the conditions of inverse EDF creation, the spatial non-uniformity of the plasma parameters of the experimental devices should be taken into consideration \cite{bib02, bib03}. Under such conditions, the EDF is nonlocal \cite{bib06}, because the unreduced kinetic equation which depends on energy and spatial variables, including the self-consistent ambipolar field, should be solved to find it.

On the one hand, this makes the procedure for solving the kinetic equation for electrons more complex. On the other hand, it increases the number of degrees of freedom of the system and results in new scenarios for the EDF formation. In particular, in the Boltzmann equation, the divergence of the spatial flow is a source (or sink) able to provide an EDF inversion.

Theoretical analysis and numerical modeling of the longitudinal structure of a classical dc glow discharge (CGD), which is the most thoroughly studied plasma object, were also conducted in \cite{bib02} and \cite{bib03}. The convenience and simplicity of its implementation and experimental study made the CGD a traditional standard for testing new ideas and diagnostics in plasma physics.

Next, we will consider the CGD at low and medium pressures, when the EDF is nonlocal. The criterion for this is \cite{bib06}:
\begin{equation}\label{eq02}
{\lambda _\varepsilon } = \lambda /\sqrt \delta   > 100\lambda  > d
\end{equation}
where $\lambda_\varepsilon$ is the electron energy relaxation length, $\lambda$ is the mean free path, $\delta  = 2m/M << 1$ is the energy exchange factor ($m$ and $M$ are the electron and atom masses, respectively) and $d$ is the characteristic length of the plasma inhomogeneity.

For atomic gases, the fulfillment of the condition \eqref{eq02} corresponds to
$p d < 5$   cm $\cdot$ Torr ($p$ is the gas pressure).

Based on the results of analytical models \cite{bib07, bib08} and numerical simulation \cite{bib02, bib03, bib09} of the longitudinal structure of the CGD with flat electrodes, the following conclusions can be made:
In a 1D short without positive column (PC), glow discharge consisting only of a cathode layer (CF) and plasma of negative glow (NG), the field reversal point (FR) is observed at the maximum of the plasma density \cite{bib07, bib08}. Under these conditions, the criterion \eqref{eq01} cannot be fulfilled. This is due to the fact that, when the density of the electrons decreases, the field decelerates them (and vice versa).

However, a longer CGD with the formation of a positive column is already two-dimensional (2D), since the existence of PC requires the presence of walls in the direction transverse to the current \cite{bib10}. Therefore, in the transition region to the PC of the Faraday dark space (FDS), a second FR point is formed, after which the field accelerates the electrons to the anode \cite{bib07, bib08}. Therefore, the criterion \eqref{eq01} is satisfied and the EDF is inverse, until the electron density continues to decrease from its maximum value in the NG \cite{bib02, bib03}.
In addition, the analysis shows that the size and position of this region depends on many factors. The latter include the ionization characteristics and range of nonlocal fast electrons that have gained their energy in the strong field of the CF, the ratio between the length $L$ and radius $R$ of the discharge tube, from the spatial distribution of the self-consistent field, the density and temperature profiles of the electrons. All these factors can reduce the region of EDF inversion. This significantly complicates experiments, since the final result is often sensitive even to minor parameter changes.

To minimize the aforementioned factors preventing creation of inverse EDF, a 2D modification of the CGD should be proposed, where  these factors act more independent of each other.
Thus, when using a hollow cathode (HC), the electrons emitted from its surface are accelerated by the cathode layer in the radial direction perpendicular to that of the discharge current \cite{bib11}. Therefore, all their ionization and the resulting plasma NG are concentrated inside the HC.
In turn, if the length $L$ of the outer region between the HC and anode is insufficient for the formation of PC ($L<\lambda_{\epsilon}$), then there is no ionization in it. Then this transition region with a weak field is an analogue of the FDS of the CGD and serves only for transporting electrons formed inside the HC to the anode. Since, in the absence of ionization, the plasma density decreases in the field accelerating electrons to the anode, condition \eqref{eq01} will be fulfilled.

In this paper, we formulate and substantiate the new approach
we are developing to solving the problem of creating an inverse electron distribution function.
Based on the kinetic criterion for the formation of an inverse EDF in an inhomogeneous plasma, the prospects of a hollow cathode glow discharge are identified. By solving the nonlocal Boltzmann equation, it is shown that in a HC discharge, the inverse EDF is realized. Based on the results of the conducted numerical modeling and theoretical analysis, practical recommendations are developed and specific conditions are identified for the experimental detection of an inverse EDF in a glow discharge with a hollow cathode.

To use the kinetic criterion \eqref{eq01}, information on the spatial distribution of the nonlocal EDF is required. Since it is applied to real experimental conditions in gas discharges with at least 2D geometry, self-consistent kinetic modeling of such regime
is an extremely resource-intensive computational task. However, there is a simpler way to estimate the fullfilling of inequality \eqref{eq01}.
Since the EDF in the low-energy region makes the main contribution to the electron density, their profiles should be similar:
$n_e\left( \mathbf{x} \right) \sim {f_0} \left( \mathbf{x}, w = 0 \right)$.
Then the condition \eqref{eq01} yields a simpler heuristic fluid criterion, identifying the areas of the most probable
implementation of the EDF inversion in various gas discharges:
\begin{equation}\label{eq03}
\nabla \varphi \cdot \nabla {n_e} < 0
\end{equation}
The criterion \eqref{eq03} depends on spatial distributions and the electron density (not on the EDF). Thus, the preliminary searches for media with inverse EDF can be conducted using some version of the fluid  model. For these models, which do not require large computational resources, many commercial software environments, tested and proven in solving various problems of plasma physics and technology, exist \cite{bib12, bib13}.

In the conducted calculations, discharge models developed in the COMSOL Multiphysics environment \cite{bib12} are used according to the previously adopted methods \cite{bib02, bib03, bib04, bib09}. The model solves the fluid balance equations for the electron density, electron energy density, heavy particle densities, and Poisson equation for determining the self-consistent electric field. The plasma-chemical model considers elastic collisions of electrons with atoms, excitation and de-excitation of a metastable level, direct and stepwise ionization, and Penning ionization.

A preliminary analysis also shows certain advantages in using activated thermionic cathodes (e.g., oxide) rather than cold ones. Their emission provides a significant determining share (almost 0.7-0.9) of the total current. This low-voltage discharge is easily ignited, and the properties of its plasma are almost similar to those of a CGD \cite{bib10}.

Next, a series of numerical experiments of glow discharges in helium at $p=1-2\,\text{Torr}$ was conducted. The discharge has the form of cylinder with
diameter $D = 2.5\,\text{cm}$; the hollow thermionic cathode has length $H = 2.5\,\text{cm}$ and is located at different distances
$L = 1 - 2.5\,{\text{cm}}$ from the flat anode.

In general, the results of the simulations are consistent with those obtained in studies on the spatial distributions of the parameters (anatomy) of a discharge with a HC \cite{bib11}. The plasma, which is weakly inhomogeneous in the longitudinal direction, is concentrated inside the HC with a diffusion (Bessel) radial charge density profile. Under the studied conditions, the maximum plasma density is located near the open end of the HC. In addition, when $L$ increases, it shifts toward the end of the HC and then slightly moves to the open part. Accordingly, the potential increases in the direction of the anode. That is, the electrons drift toward the anode in a longitudinal extracting field.

Typical simulation results for a discharge in helium at $p = 2\,\text{Torr}$ and the open part length $L = 1.5\,\text{cm}$
are presented in the sequel.

Figure 1 shows 2D distributions of the main parameters of interest for the open part of the discharge for $z > 25\,{\text{mm}}$.
\begin{figure}
\includegraphics[width=13cm]{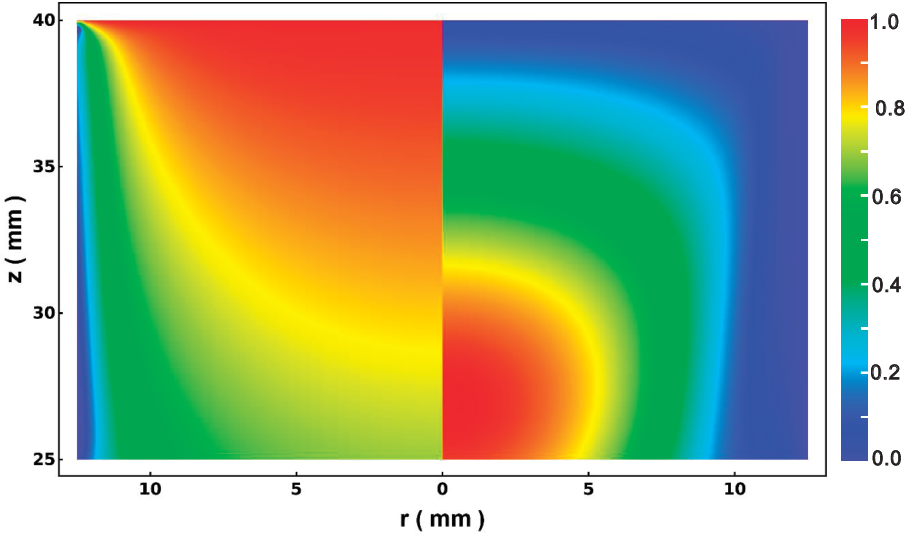}
\caption{\label{fig:fig1}
2D profiles of electron density and electric potential.
}
\end{figure}
It can be seen that, in the direction towards the anode, after the maximum electron density, the latter decreases with the increase of the electric potential $\varphi$. That is, the conditions of EDF inversion are fulfilled according to the fluid criterion \eqref{eq03}.

For a detailed analysis, we will further consider the results of simulations with the solution of the nonlocal Boltzmann kinetic equation for the isotropic component  $f_0$ \cite{bib02, bib03}:
\begin{equation}\label{eq04}
\nabla \cdot \mathbf{\Phi} +\frac{\partial\Gamma}{\partial w} = S^{*}\left( f_0 \right),
\end{equation}
\begin{equation}\label{eq05}
\mathbf{\Phi} = \frac{\gamma}{3} \sqrt{w}\,{\mathbf{f}}_1 ,\,\,
\Gamma = -\mathbf{E}\cdot\mathbf{\Phi} -
w^{3/2} \delta \nu_a \left(T\,\frac{\partial f_0}{\partial w} + f_0 \right) ,
\end{equation}
where ${\mathbf{\Phi }}$ and $\Gamma $ are respectively the spatial and energy phase fluxes, $\gamma  = \sqrt {2e/m\,} $,
$\mathbf{f}_1=-\lambda\left(\nabla f_0 - \mathbf{E}\,{\partial f_0}/{\partial w}\right)$ is the directed component of EDF,
${S^*}\left( {{f_0}} \right)$ is the integral of inelastic collisions,
${\nu_a}$ is the frequency of elastic scattering and $T$ is the gas temperature.
The details of the kinetic equation \eqref{eq04}, boundary conditions, and solution methods are presented in \cite{bib02, bib03, bib09}.

Figure 2 shows $f_0(z, r=0, w)$ for different values of the axial coordinate $z$.
\begin{figure}
\includegraphics[width=13cm]{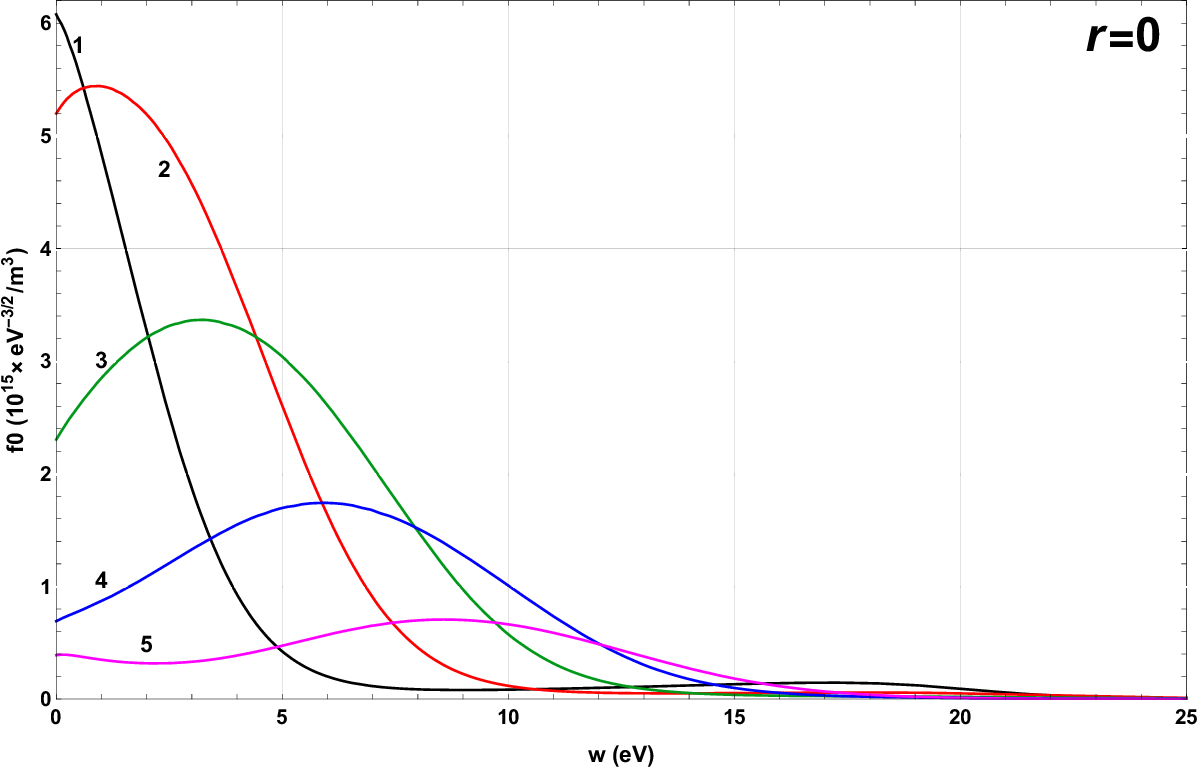}
\caption{\label{fig:fig2}
${f_0}\left( {z,w} \right)$ for $z_1$ = 20 mm (1), $z_2$ = 24 mm (2), $z_3$ = 28 mm (3), $z_4$ = 32 mm (4),
and $z_5$ = 36 mm (5) on the discharge axis ($r = 0$).
}
\end{figure}
As expected, in the region of decreasing electron density $z > 25\,\text{mm}$, where the conditions \eqref{eq01}, \eqref{eq03} are satisfied, an inversion of the EDF in the region of low energies can be observed. In this case, as the anode is approached, the width of this region in the energy scale increases with the decrease the electron density.

To analyze and explain the obtained result, we recall that under these conditions, condition \eqref{eq02} of nonlocality is satisfied. Therefore the total energy
$\varepsilon = w - e \varphi$  \cite{bib06} which is an {\it integral of motion} is the argument of the EDF. This means that different groups of electrons reach the anode preserving its total energy and do not “mix” due to collisions.
Fig. 3 shows the EDF from Fig. 2 as a function of the total energy $\varepsilon$.
\begin{figure}
\includegraphics[width=13cm]{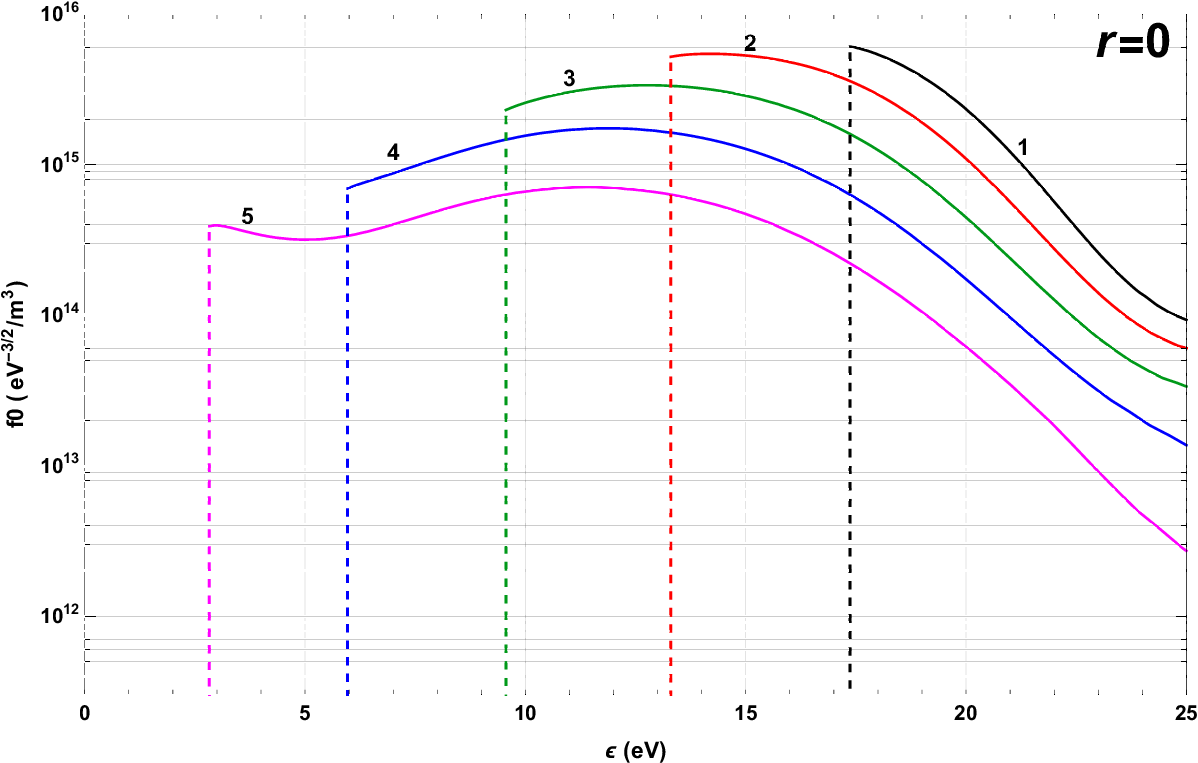}
\caption{\label{fig:fig3}
EDF as a function of the total energy $\varepsilon$.
The notations are the same as in Fig. 2. The dotted vertical lines correspond to $w=0$.
}
\end{figure}

It can be seen from Figure 3 that the inversion is for all the EDFs at a total energy ${\varepsilon _2} \leqslant 13\,{\text{eV}}$, for $z > z_2$.
Similar to Figure 2, when approaching the anode, the width of this region increases and the electron density decreases.
However, the EDF inversion starts with approximately the same ${\varepsilon_2}$.
In other words, $\left( {{z_2},{\varepsilon_2}} \right)$ on the phase plane $\left( {z,\varepsilon } \right)$ is a boundary point. From it, the EDFs start to decrease toward higher and lower energies.
In the total energy scale, the EDFs in Figure 3 have almost the same behavior. However, the considered case is different from the classical Holstein-Tsendin model of complete nonlocality \cite{bib06} which is realized for the trapped electrons. In this case, after averaging over the region of phase space accessible to electrons, the EDF becomes a function of only one variable $\varepsilon $. In the considered case, due to the departure to the anode, the absolute values of the EDF decrease and a small (of the order of $L/{\lambda _\varepsilon } < \,\,1$) deformation of the energy dependence of the EDF occurs. Therefore, the EDFs of these electrons are always function of two variables ($\left( {z, w} \right)$ or $\left( {z, \varepsilon } \right)$ ).

Figure 4 shows a drawing explaining the formation of the inversion of the nonlocal EDF in the open part of the HC discharge. This drawing is constructed using the aforementioned simulation data. The right part of the figure shows the calculated dependence of the potential $e\varphi \left( z \right)$on the point ${z_1}$ in Figure 2, corresponding to the maximum plasma density. Point $z = L$ is the anode, which potential is set to zero. Since the kinetic energy of the electron $w = 0$ on the curve of the potential dependence $\varepsilon  = e\varphi \left( z \right)$, there are no electrons below this curve (darkened one). At $L < {\lambda _\varepsilon }$,  nonlocal electrons move to the anode on the phase plane along the horizontal straight lines (cf. black arrows in Figure 4) while their total energy $\varepsilon $ is conserved.
The initial value of the potential $e\varphi \left( {{z_1}} \right)$ (black dotted horizontal line) divides the phase plane $\left( {z,\varepsilon } \right)$ into two regions: the upper I where
$\varepsilon  > e\varphi \left( {{z_1}} \right)$ and ionization sources from the HC exist, and the lower II where the generation of charges does not occur. In the absence of ionization in region II, only electrons from zone I can enter due to their deceleration in elastic collisions. Therefore, in the nonlocal mode ($L < {\lambda _\varepsilon }$), the maximum of the EDF will be near the total energy $\varepsilon  = e\varphi ({z_1})$, so that further to the anode, it will decrease towards higher and lower energies. This indicates that an inverse EDF will be observed.
The left part of Figure 4 shows the EDF dependencies calculated in this study (cf. Figure 2 and 3), which confirms the aforementioned conclusions.
\begin{figure}
\includegraphics[width=13cm]{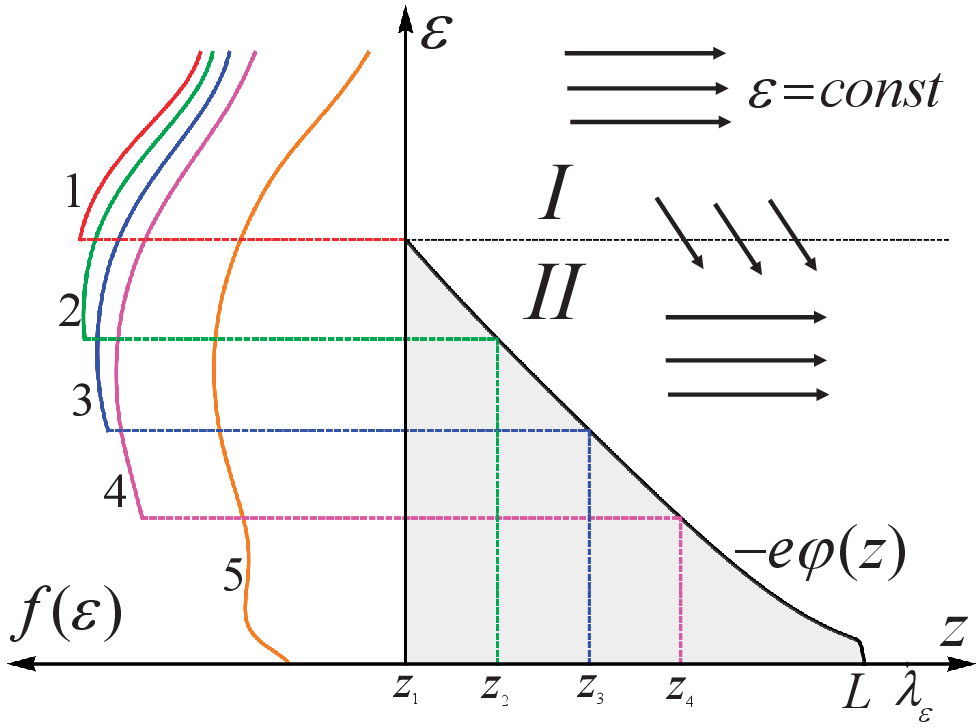}
\caption{\label{fig:fig4}
Drawing explaining the formation of the EDF inversion in the open part of the HC discharge.
}
\end{figure}

This consideration can also be explained using the following analytical model.
In region II for the kinetic energies below the helium excitation threshold of 19.8 eV, the kinetic equation \eqref{eq04} can be written as:
\begin{equation}\label{eq06}
\frac{\partial }{{\partial z}}\left( {D\frac{{\partial {f_0}}}{{\partial z}}} \right) + \frac{\partial }{{\partial \varepsilon }}\left( {V\,{f_0}} \right) = 0,
\end{equation}
 where $D = \gamma w \lambda /3$ and $V = {w^{3/2}}\delta \,{\nu_a}$ are the spatial diffusion and “friction force” coefficients, respectively. With a simplifying assumption that the coefficients $D$ and $V$ are constants (at some vicinity of the $\varepsilon  = {\varepsilon _0}$ line on the phase plane), Eq. \eqref{eq06} is reduced by the replacing of variable $t = {\varepsilon_0} - \varepsilon $ to the thermal conductivity (diffusion) equation, which has the following known fundamental solution:
\begin{equation}\label{eq07}
g\left(z,t\right) = {\left( {4\pi \left( {D/V} \right)t} \right)^{ - 1/2}}\exp \left( { - V\,{z^2}/\left( {4D\,t} \right)} \right).
\end{equation}
Thus, the function $G(z, \varepsilon)=g(z-z_0, \varepsilon_0-\varepsilon)$ is a solution of Eq. \eqref{eq06}
with the initial condition $G(z, \varepsilon_0)=\delta(z-z_0)$ in the $\varepsilon  < {\varepsilon _0}$ region.
This solution decreases with the distance from the $\varepsilon=\varepsilon_0$ line on the phase plane $(z, \varepsilon)$.
In the considered situation, electrons can enter zone II only from zone I  (cf. Figure 4). Accordingly, the solution of
kinetic equation \eqref{eq06} in zone II is a superposition of the solutions $g(z-z_0, \varepsilon_0-\varepsilon)$ for all points $z_0$
of the boundary $\varepsilon=\varepsilon_0$:
\begin{equation}\label{eq08}
{f_0}\left( {z,\,\varepsilon } \right) = \int\limits_{z_1}^L {{f_0}\left(z_0, \varepsilon_0 \right) g\left( z - z_0, \varepsilon_0 -\varepsilon  \right)d{z_0}} .
\end{equation}
Thus,  solution \eqref{eq08} exponentially decreases with the distance from the boundary $\varepsilon  = {\varepsilon _0}$.
Consequently, in zone II, ${f_0}\left( {z,\varepsilon } \right)$ increases with the increase of $\varepsilon $, which explains the inversion of the EDF in this zone.

\section{Conclusion}
This paper presents an auxiliary approximate fluid criterion based on the kinetic criterion for the formation of an inverse EDF in a non-uniform inhomogeneous plasma. It allows for a rapid search for potential plasma media using a standard generally-accepted fluid model.
The conducted theoretical analysis and numerical modeling show the fundamental possibility of creating an inverse EDF in the near-anode nonlocal plasma of a hollow-cathode glow discharge.

By solving the complete Boltzmann kinetic equation in energy and coordinate variables, inverse EDFs are determined, and the physical mechanisms of their formation under the studied conditions are analyzed.

Based on the results of the numerical modeling and theoretical analysis, practical recommendations are developed and specific conditions are identified for the experimental detection of an inverse EDF in a glow discharge with a hollow cathode.

In summary, the following procedure for searching for inverse EDFs is recommended:

1) The objects of research are spatially inhomogeneous media with a nonlocal EDF, when the electron energy relaxation length is greater than the characteristic size of the discharge region of interest (i.e., ${\lambda _\varepsilon } > L$). For atomic gases, this corresponds to $pL < 5\,{\text{cm}} \cdot {\text{Torr}}$.
When searching for inverse EDFs, attention should be first paid to the near-anode plasma of hollow cathode discharges, to the inhomogeneous positive column of the discharge
(e.g., created by introducing a diaphragm), and to the two-chamber designs of ICP and CCP discharges, where plasma is generated in the first chamber and diffuses into the second, ballast chamber.

2) Based on the results of the heuristic analysis, a set of potential media with possible EDF inversion is taken into consideration. For these purposes, fluid simulations are conducted. In addition, based on the determined field and electron density distributions, the regions of fulfillment of the auxiliary approximate fluid criterion are identified, when the electron density decreases in the field accelerating electrons to the anode (i.e., positive electrode).

3) For the final solution in the regions of potential inversion of the EDF, the complete kinetic equation is solved in the variables of energy and coordinates, the conditions of the inverse EDF are determined, and recommendations for experimental search are provided.

Thus, in this work, based on the proposed novel approach for creating an inverse electron distribution function, practical recommendations for the experimental detection of the inverse EDF with an indication of specific conditions in a glow discharge with a hollow cathode are developed.
\begin{acknowledgments}
This work is supported by National Natural Science Foundation of China (NSFC, Contracts No.12205067, 12175050), Natural Science Foundation of Heilongjiang Province of China (YQ2024A008).
\end{acknowledgments}

\end{document}